\documentclass{PoS}

\usepackage{graphicx}
\usepackage{amsmath}

\newcommand{\bee}{\begin{equation}}
\newcommand{\ee}{\end{equation}}
\newcommand{\beea}{\begin{eqnarray}}
\newcommand{\eea}{\end{eqnarray}}

\def\Tr{{\rm Tr}}

\title{The one-flavor quark condensate and related problems}

\ShortTitle{The one-flavor quark condensate and related problems}

\author{\speaker{Thomas DeGrand}\\
        University of Colorado\\
        E-mail: \email{degrand@pizero.colorado.edu}}

\abstract{
I describe a recent calculation (by me, Hoffmann, Liu, and Schaefer)
of the chiral condensate in one-flavor QCD using
numerical simulations with overlap  fermions. The condensate
is extracted by fitting the distribution of low lying eigenmodes of the Dirac operator
in sectors of fixed topological charge to the predictions of Random Matrix Theory.
Our results are in excellent agreement with estimates from the orientifold large-N
expansion. Much interesting physics surrounds this calculation, which I will  highlight.
}

\FullConference{XXIVth International Symposium on Lattice Field Theory\\
                July 23-28, 2006\\
                Tucson, Arizona, USA}

\begin{document}

Theories like QCD, but with small differences, can teach us things about QCD. QCD with one flavor
of dynamical fermion is such a theory: it is related, by a remarkable 
path, to the $N_c\rightarrow \infty$ limit of  ${\cal N}=1$
supersymmetric Yang-Mills theory, especially to one
of its exactly-known observables, the gluino condensate. 

We lattice people are not the only ones interested in strongly coupled gauge field theories.
Supersymmetry is a powerful tool for understanding these systems, and it is an important
and active area of research, to extend results from supersymmetric theories to
 non-supersymmetric ones.
One way to do this is to replace the degrees
of freedom in supersymmetric field theories with new ones, while still preserving
desirable features. The large-$N_c$ (number of colors) limit is an important part of this program.
(For an early attempt, see \cite{Strassler:2001fs}.)

\begin{figure}
\begin{center}
\includegraphics[width=0.4\textwidth,clip]{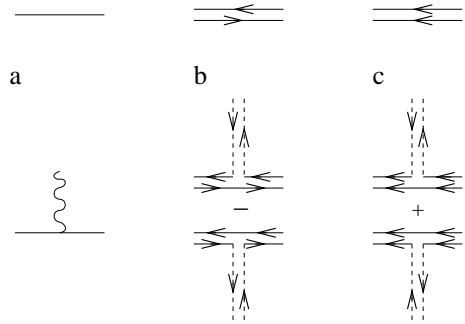}
\end{center}
\caption{(a) The fermion  propagator and the fermion--fermion--gluon
vertex. (b) ${\cal N}=1$
SYM theory. (c) Orientifold daughter}
\label{fig:fey1}
\end{figure}

Recently, Armoni, Shifman, and
Veneziano~\cite{Armoni:2003fb,Armoni:2003gp,Armoni:2003yv,Armoni:2004ub} (ASV)
 suggested a new large-$N_c$ expansion
with some remarkable features.
In contrast to the 't Hooft large-$N_c$ limit~\cite{'tHooft:1974hx}
 ($N_c\rightarrow \infty$, $g^2 N_c$ and $N_f$ fixed,
with quarks in the fundamental representation of $SU(N_c)$), quarks are placed in the
two-index antisymmetric representation of $SU(N_c)$. Now in the
$N_c\rightarrow \infty$, $g^2 N_c$ and $N_f$ fixed limit of QCD, quark effects are not
decoupled, because there are as many quark degrees of freedom as gluonic ones, $O(N_c^2)$
in either case. In Ref.~\cite{Armoni:2004ub} the
authors have argued that a bosonic sector of ${\cal N}=1$ super-Yang-Mills (SYM) theory
is equivalent to this theory in the large-$N_c$ limit. ${\cal N}=1$ SYM is a theory
of adjoint gluons and their gluino (Majorana fermion) partners, and the equivalence
of these theories in perturbation theory can be seen by comparing the vertices,
as in Fig. \ref{fig:fey1},
taken from Ref. \cite{Armoni:2004uu}.  The large-$N_c$ QCD-like theory
is called ``orientifold QCD.''

The perturbative connection of orientifold QCD to ${\cal N}=1$ SYM is uncontroversial.
In Ref.~\cite{Armoni:2004ub},
ASV have presented a nonperturbative proof of the connection.
This proof  has been extended by Patella~\cite{Patella:2005vx}
 to lattice regularized theories. Recently, Yaffe and \"Unsal~\cite{Unsal:2006pj}
have argued that the proof of ASV is incomplete: that orientifold QCD
and  ${\cal N}=1$ SYM
have a different phase structure
on spacetimes with small compact dimensions, in which charge conjugation symmetry
 is spontaneously broken. Only when the two theories have
identical vacua can the proof hold.

I am certainly not competent to comment more on this subject, so let us see what
orientifold QCD might have to do with a lattice project:
For $N_c=3$, orientifold QCD is equivalent to QCD with a single quark flavor
in the fundamental representation of $SU(3)$. This equivalence can be seen in the 
first and second  terms in the $\beta$ function and in the lowest order
anomalous dimension for the running quark mass (or quark condensate), as Table \ref{tab:TTT}, taken from Ref.
\cite{Armoni:2004uu} shows.
This means that
if the proof of nonperturbative equivalence is correct,
 nonperturbative quantities (in the bosonic sector) computed in super-Yang-Mills theory
can be related to corresponding ones in one-flavor QCD, up to $1/N_c$ effects.

The analog of the quark condensate in ordinary QCD is the gluino condensate in 
 ${\cal N}=1$ SYM.
It can be calculated exactly in large-$N_c$ using saddle point methods (Ref.
 \cite{Davies:1999uw} is a recent reference with a complete citation path).
ASV used this exact result to estimate~\cite{Armoni:2003yv} the quark condensate
in one-flavor QCD from the value of the gluino condensate in SYM.
They found (with our sign conventions)
\bee
\Sigma = \{0.014, \ 0.021, \ 0.028\} \ {\rm GeV}^3
\label{eq:asv}
\ee
in the $\overline{MS}$ scheme at $\mu=2$ GeV. The spread of values gives their estimate
of $1/N_c$ corrections (basically $1\pm 1/3$).

Now we come to the lattice: 
Last November, Schaefer and I had most of the
solution to the problem of how to do Hybrid Monte Carlo for any $N_f$,
 using overlap fermions\cite{DeGrand:2006ws}.
But what to do with it? I remembered the ASV prediction.
Over Christmas vacation we started simulations to check this number
and the preprint \cite{DeGrand:2006uy} came out in May. We hit their number bang on!

We have to deal with a certain amount of imprecise language related to the condensate:
The quantity $\langle \overline q q \rangle$ is one definition of the condensate.
Rather than measuring it directly, we will determine
the  particular combination of the coefficients of the low energy effective field theory,
 $\Sigma= f^2 B$, in the usual parameterization for $N_f >1$ QCD,
\bee
{\cal L}_2 =
\frac{f^2}{4}\Tr( \partial_\mu U \partial_\mu U^\dagger)+ B \frac{f^2}{2}\Tr[ M(U+U^\dagger)].
\ee
One expects that the quantity $\langle \overline q q\rangle$ (as computed, for
example, in a lattice simulation at some quark mass $m_q$ and simulation volume $V$) is a
function of $\Sigma$, $f$, $m_q$, and simulation volume $V$. $\Sigma$ and $f$ are the
 interesting quantities, and a direct lattice measurement of $\langle \overline q q\rangle$
from several quark masses
would have to be converted to a prediction of $\Sigma$ and $f$, by fitting  it
to the appropriate functional form from chiral perturbation theory. The same thing would have to
be done if one measured observables, like the pseudoscalar mass and $f_\pi$,
and used the GMOR relation to infer $\Sigma$.

$N_f=1$ QCD is a peculiar theory. Chiral symmetry is anomalous.
There are no Goldstone bosons, just the eta-prime, which
gets its mass through the anomaly. 
The $\Sigma$ which we are about to extract is therefore not an order parameter
of spontaneous chiral symmetry breaking. However, there still exists a well defined
low-energy description of $N_f=1$ QCD.
It has been given by Leutwyler and Smilga~\cite{Leutwyler:1992yt} to which we refer
the reader for details.
They show that up to terms of order $m^2V$  the partition function is
\bee
Z=\exp\big\{\Sigma V  {\rm Re}(me^{-i\theta}) \big\}
\ee
with $\theta$ the vacuum angle.
$\Sigma$ is the infinite volume zero quark mass limit of $-\langle \bar q q \rangle$
at $\theta=0$.

\begin{table}
\begin{center}
\begin{tabular}{|c|c|c|c|}
\hline
$\frac{\mbox{ Theory} \to  }{\mbox{  Coeff} \downarrow} $ \rule{0cm}{0.5cm} & 1f-QCD &  Orienti A  & S
YM \\
\hline\hline
\rule{0cm}{0.7cm}$\beta_0$  & $\frac{11}{3}\, N-\frac{2}{3}$&$3N+\frac{4}{3}$ &$3N$\\
\hline
\rule{0cm}{0.7cm}$\beta_1$  & $\frac{17}{3}N^2-\frac{13}{6}N+\frac{1}{2N}$&
$3N^2 +\frac{19}{3}N-\frac{4}{N}$ &$3N^2$\\
\hline
\rule{0cm}{0.7cm}$\gamma$ & $\frac{3(N^2-1)}{2N}$&$\frac{3(N-2)(N+1)}{N}$ &$3N$\\
\hline
\end{tabular}
\end{center}
\caption{A comparison of the  beta functions and anomalous dimensions for the three theories
\label{tab:TTT}}
\end{table}

There is a reasonably inexpensive way to directly compute $\Sigma$:
it involves measuring the distribution of the low eigenmodes of the Dirac operator, in sectors of
 fixed topology
$\nu$ in a simulation volume $V$.
The probability distribution of individual eigenvalues $\lambda_n$ is given by
Random Matrix Theory (RMT)~\cite{Shuryak:1992pi,Verbaarschot:1993pm,Verbaarschot:1994qf}
as a function of the dimensionless quantity $\lambda_n \Sigma V$,
which depends parametrically
on the combination $m_q \Sigma V$ and, of course, $N_f$.
We use the specific method and predictions from
 Refs.~\cite{Damgaard:2000qt,Damgaard:2000ah}.

The connection is only supposed to work deep in the $\epsilon$ regime of QCD, but the
eigenmode distribution is very robust and this does not seem 
to be a necessary constraint in practice.

So the calculation involves several parts:

\begin{itemize}
\item
Perform simulations in sectors of fixed topology
\item
Measure the eigenmodes and fit their distributions to the RMT expression: this gives 
the lattice-regulated $a^3\Sigma$
\item
Determine $a$: we used the potential and the Sommer parameter \cite{Sommer:1993ce}
\item
Convert to $\overline{MS}$: we used the RI-MOM method \cite{Martinelli:1994ty}
\end{itemize}

How we did the simulations \cite{DeGrand:2006ws} is worth a paragraph.
The last two items are pretty straightforward. So is computing the eigenmodes.
Fitting the distributions is also a little nonstandard, so I'll fill that in, too.

When one is dealing with chiral symmetry on the lattice, it is very convenient to work with a 
fermion action which is chiral. That way, the physics of spontaneous symmetry breaking 
(including the anomaly) is not masked by explicit chiral symmetry breaking from the
lattice discretization. While standard lore says that it is in principle possible to correct for
chiral symmetry breaking effects in the bare action during the analysis, in practice
this can be difficult and there is always the danger of new effects which one did not
plan on (like exceptional configurations for Wilson-type actions). Why go looking for trouble?

Thus we are led to use overlap~\cite{Neuberger:1997fp,Neuberger:1998my}
 fermions, which exactly encode
chiral symmetry through the Ginsparg-Wilson~\cite{Ginsparg:1981bj}
 relation. It happens that, as an extra treat, it is
possible to simulate any number of flavors of overlap fermions, without requiring
any degeneracy in the quark mass spectrum, using the exact Hybrid Monte Carlo algorithm.

Due to the Ginsparg-Wilson relation, overlap fermions have the nice properties that
the spectrum of the squared massless Dirac operator $H^2=D^\dagger D$ commutes with $\gamma_5$
and has degenerate opposite-chirality eigenfunctions, apart from chiral zero modes and their
partners at $H^2=4R_0^2$ (where $R_0$ is the radius of the Ginsparg-Wilson circle).
The corresponding eigenmodes of $H=\gamma_5 D$ and $D$ itself can be found by diagonalizing
the $2\times 2$ degenerate subspaces of $H^2$. Then \cite{Bode:1999dd,Cundy:2005mr}
the contribution of the paired
modes to the determinant of a single flavor of overlap
fermions is given by $\det H_\sigma^2$, the determinant of $H^2$ evaluated in a single
chirality sector. 
It is included in the action with a single chiral pseudofermion for each flavor,
$\Delta S =  \phi_\sigma H_\sigma(m)^{-2}\phi_\sigma$.
The contribution of the zero modes can be included in the HMC by
direct addition to the action at a topological boundary (the extra factor is $N_f|Q|\ln(m_q/(2R_0))$
for $N_f$ degenerate flavors). For analysis, there is a different
weighting of $\nu=0$ and $\nu \ne 0$ configurations.

What was never written down (maybe it is trivial) is how to
initialize the pseudofermions:
we need to generate random numbers to initialize $\phi_\sigma$.
The trick is to begin with a chiral random source $R_\sigma$ and to
use the  Zolotarev formula to construct $\phi_\sigma = \sqrt{H_\sigma(m)^2} R_\sigma$.

We used the algorithm of
Ref. \cite{Fodor:2003bh}. In it, one must monitor the eigenmodes of the ``kernel
operator''
$h(-R_0)$ in $D=R_0(1+\gamma_5 \epsilon(h(-R0)))$. When it develops a zero mode,
the topology of the underlying gauge configuration changes and with it, there is a
step
discontinuity in the fermionic action.
Random matrix theory wants eigenmodes in sectors of fixed topology. We generate those
simply
by forbidding tunneling events in the molecular dynamics evolution, and evolving
in sectors
of fixed $\nu$.

To do the calculation, we collected data on a few P4's and P4E's for a few months, 
on $10^4$ and $8^3\times 12$ lattices, at a lattice spacing
of about 0.15 fm..
(The  $10^4$ data set had about 500 trajectories each for winding number $\nu=0$ and 1.)

Dynamical overlap is not cheap but it is completely feasible for small projects as long
 as one is willing to be creative (or maybe Baroque?) \cite{DeGrand:2004nq,DeGrand:2005vb}. 
The essential ingredient is a fat link gauge connection.
We used stout links \cite{Morningstar:2003gk},
 three steps with $\rho=0.15$. As far as the overlap goes, more smearing is better
and 3 times 0.15 is about 2.5 times faster than the 2 times 0.15 of our previous
 work~\cite{DeGrand:2004nq,DeGrand:2005vb}.
The improvement comes from decoupling the fermions from UV gauge fluctuations
which would generate small eigenmodes of $h(-R_0)$. This decreases the
condition number of $h(-R_0)$ and speeds up the calculation of $D$.
 The dark side (if there is one) of a fat link
action is that the fat links make the action more spread out than a conventional thin
link action. Minimizing this spread is part of ``action engineering,'' like minimizing the
range of the fermionic couplings. We have performed the usual tests of locality on our action
and never seen anything peculiar. Remember, thin links and fat links are just choices
for the bare action which differ (in a Symanzik sense) through irrelevant operators.
Formally, thin and fat link actions
 are both in the universality class of QCD. We are allowed to tune
irrelevant operators as we please to ease the computational burden while
preserving symmetries -- not to do so
is bad software engineering.

If one wants to use eigenmodes, a kernel which looks ``overlap-like'' is also essential,
otherwise the eigenmode part of the code is prohibitively expensive. (I have never been able to do
anything with thin link Wilson-kernel overlap; it is too expensive for work stations.
The problem is in the eigensolver: one begins with a set of trial eigenmodes
which are then iteratively improved. Without some good idea, the beginning modes
are typically random vectors. Improving them takes many iterations, which when
done with the overlap action is very slow.
One can gain a lot of time using better eigenmodes, from some ``overlap-like''
action, as an intermediate step.
 The overlap actions I use are built on  ``overlap-like'' kernels, but
the overlap action with a Wilson kernel has a very different spectrum from a Wilson action.)

All of this is well documented for quenched simulations \cite{DeGrand:2000tf}. 
I can't help thinking that there are more tricks out there.
Oh, for another factor of five speedup...

So we collected a set of fermionic eigenmodes. We want to fit their probability
distributions to RMT formulas.
The analysis is a little nonstandard, because the eigenmodes are continuously distributed.
A referee led us to the Kolmogorov-Smirnov test~\cite{Press:2002} 
as a measure for the goodness of the
fit. It compares the cumulative distribution function of the data
$C(x)$ to
the theoretical prediction $P(x)=\int_{-\infty}^x f(x) {\rm d} x$.
$C(x)$ is the fraction of eigenvalues  with a value smaller
than $x$.

\begin{figure}
\begin{center}
\includegraphics[width=0.4\textwidth,clip,angle=-90]{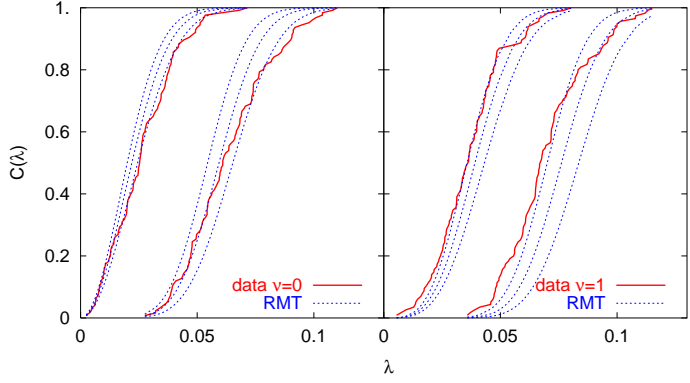}
\end{center}
\caption{Cumulative  distributions from RMT fit to eigenmode distributions. For each
mode, the three dotted lines, running from left to right,
 correspond to fits to  the lowest mode  in the $\nu=1$  sector only,
 the lowest mode $\nu=0$ and $|\nu|=1$, and the lowest mode in $\nu=0$ only.
\label{fig:RMTc1}
}
\end{figure}

The quantity of interest is the largest deviation of $P$ and $C$:
$D=\max_x |P(x)-C(x)|$. From this the confidence level is given by
\bee
Q_{KS}\left((\sqrt{N}+0.12+0.11/\sqrt{N})D\right)
\ee
with
\bee
Q_{KS}(\lambda)= 2\sum_{j=1}^\infty (-)^{j-1}\exp(-2j^2\lambda^2).
\ee
In fits to a single eigenmode distribution we maximize this quantity.
When fitting to more than one mode, we maximize the product over the individual
confidence levels.
 The errors on the fit parameter $\Sigma$ are determined by the
bootstrap procedure. An example of such a fit is shown in Fig. \ref{fig:RMTc1}.
After a lot of angst (which modes to fit, what about correlations...) $\Sigma a^3$
turned out to be remarkably robust: it didn't matter what we did.

Completing the calculation with the lattice spacing from the Sommer parameter and
the matching factor from RI-MOM, we
found
\beea
r_0^3 \Sigma(\overline{MS},\mu=2 \ {\rm GeV})& = &
 Z_s(\mu,a) \times  \Sigma a^3 \times (\frac{r_0}{a})^3 \nonumber \\
 &=& (0.86(3))\times(0.0096(3)) \times(3.37(10))^3 \nonumber \\
&=& 0.317(22). 
\eea
With $r_0=0.5$ fm, this is
\bee
\Sigma(\overline{MS},\mu=2 \ {\rm GeV}) = 0.0194(20){\rm GeV}^3 
\ee
which agrees pretty nicely with Eq. \ref{eq:asv}.

The summary of McNeile \cite{McNeile:2005pd} shows that the $N_c=3$ condensate is not
very $N_f$ dependent. Indeed, Schaefer, Liu and I just finished \cite{DeGrand:2006nv}
 an $N_f=2$ measurement
using basically identical techniques to what I have described here, and
we find $\Sigma = 0.0225(25)$ GeV${}^3$
=(282(10) MeV)${}^3$.
There is actually an annoying systematic in this number: In finite volume,
there is a first order correction to the condensate, basically  the
one loop graph from Goldstone bosons which are emitted from the propagating
pseudoscalar and which are absorbed at an image point of the vertex.
The correction is  $\Sigma \rightarrow \rho \Sigma$,
where
\bee
\rho = 1 + \frac{N_f^2-1}{N_f}\frac{c(l_i/l)}{f_\pi^2L^2}
\label{eq:rho}
\ee
and $c(l_i/l)$ depends on the geometry\cite{Gasser:1986vb}. 
This term is absent in $N_f=1$ because there are no Goldstones.
We haven't measured $f_\pi$, but with
93 MeV, in our volume for $N_f=2$,
 $\rho \sim 1.43$. Fortunately, people publish $\Sigma^{1/3}$, not $\Sigma$!

After the paper came out, Veneziano reminded me that their prediction was really
for $\Sigma/\Lambda^3$ ``without going through actual
experimental numbers.''
 For us lattice people, a ratio of nonperturbative quantities with no
(or minimal) intrusion of perturbation theory is much cleaner than $\Sigma/\Lambda^3$,
and the definition of $\Lambda$ is exquisitely sensitive to a determination of a coupling:
recall that
\bee
a \Lambda = \left(\frac{16\pi^2}{\beta_0g^2(a)}\right)^{\beta_1 /( 2 \beta_0^2 )}
   \exp\left(- \frac{8\pi^2}{\beta_0 g^2(a)}\right) .
   \label{eq:2.39}
\ee
Tadpole-ology (plaquette $\rightarrow \alpha_V(q^*)\rightarrow \Lambda$) at
one value of the lattice spacing
is just too unstable
to be useful. However, the ALPHA collaboration has used the Schr\"odinger functional real space
renormalization group to compute  \cite{Capitani:1998mq,DellaMorte:2004bc}
 the quantities
\beea
\Lambda(\overline{MS})r_0 &=& 0.62(4)(4) ; \qquad N_f=2 \nonumber \\
                          &=& 0.60(8)  .\qquad N_f=0 \nonumber \\
\label{eq:alpha}
\eea
If we take $\Lambda(\overline{MS})r_0=0.61(6)$ for $N_f=1$, then
\bee
\frac{\Sigma}{\Lambda^3} = 1.4(4)
\ee
while ASV want 0.6 to 1.1 for the ratio. This is not so bad! (The difference between this and
Eq. \ref{eq:asv} is a 15 per cent shift between their $\Lambda$ and the
interpolated ALPHA value.)

Armoni, Shore, and Veneziano have also predicted $\Sigma$ for other $N_f$'s \cite{Armoni:2005wt}.
They add $N_f-1$ fundamental flavors to the mix, and get
\bee
\frac{\Sigma}{\Lambda^3} = \frac{1}{4\pi^2}(8-N_f)
\label{eq:asvnf}
\ee
for the RGI condensate. They need a coupling constant to convert this to an
$\overline{ MS}$
number.
Their publication only presents a band, since when they do the conversion
from RGI to $\overline{MS}$ they consider a range of coupling constants.
 However, taking  the $N_f=2$ value of $\Lambda(\overline{MS})$ from
 Eq.~\ref{eq:alpha} and inverting it to give a coupling constant,
 $\Sigma^{1/3} = 247$ MeV. More accurate
lattice measurements of $\Sigma$ (and $\Lambda$) vs $N_f$ could test
the $N_f$ dependence in Eq.~\ref{eq:asvnf}.

I am not sure what to do next. Unfortunately, the only nonperturbative quantity
which can be computed in SYM is the gluino condensate 
(as far as I can tell, from asking many people), although
to compensate, it is supposedly exact.  So other predictions typically involve ratios of masses
without some connection to the condensate:
\bee
\frac{m^2_{\eta'}}{m_\sigma^2} = 1 + O(1/N_c)
\ee
(From Ref. \cite{Armoni:2003fb};
the $O(1/N_c)$ corrections have been computed by Sannino and Shifman \cite{Sannino:2003xe}),
and a similar degeneracy for hybrids \cite{Gorsky:2004nb}. An $N_f=1$ meson is just like
a flavor singlet meson in ordinary multiflavor QCD, so both the $\eta'$ and $\sigma$
have disconnected (``hairpin'') contributions. These are difficult and noisy. In addition, the
scalar operator has a vacuum expectation value, so its signal is like a scalar
glueball's: the exponential which gives the mass dives under a constant background.
 I have successfully avoided trying to do these for four months, now.

$N_f=1$ QCD has other intriguing properties: 
At negative quark mass, $N_f=1$ QCD  may have a phase in which CP is broken.
(This can occur for any $N_f$ if the quark mass matrix has positive and negative
eigenvalues. This observation goes back to Dashen~\cite{Dashen:1970et}; 
see Smilga~\cite{Smilga:1998dh} and Creutz~\cite{Creutz:2000bs} for recent discussions.)
It is possible to simulate QCD at $m_q<0$ (or even complex mass) with overlap fermions
by reweighting a real-mass simulation. The relevant derivation has been given by
D\"urr and Hoelbling~\cite{Durr:2006ze}, who briefly studied the Schwinger model.
Complex mass is equivalent to a theta vacuum, another unvisited area of QCD for
lattice simulators.

``The purpose of computing is insight, not numbers'' (Hamming), so did we learn anything?
It is always easiest to say No. The only quantitative prediction from ${\cal N}=1$ SYM
is the condensate, and it has unknown $1/N_c$ corrections. The spectrum of QCD depends
on $N_f$ in an uncontrolled way. The only way to make predictions relevant to the real world
is with simulations with
three flavors of light quarks, all in the chiral regime. Even then, one must
simulate at physical quark masses, unless the observable being studied has a well-behaved
expansion in chiral perturbation theory allowing one to extrapolate in quark masses.

And yet--

When we teach about the spectrum of hydrogen in an introductory quantum mechanics class,
the story is not ``we do the calculation and the answer is 13.6 eV.''
There is systematic expansion (in $\alpha$ or equivalently in $v/c$) which allow us to
make successively more accurate predictions.
The parameters we have at our disposal in QCD are $N_c$, $N_f$, the color representations
of the quarks, and the quark masses.
Perhaps the zeroth order QCD calculation (like $V(r)=-e^2/r$ for hydrogen) is some extreme
value of one or all of these parameters. Lattice tests of QCD-like theories might
tell us new things about QCD, if they could validate extrapolations
of analytic results from those theories. We won't know if we don't try!

And this test worked: ASV successfully predicted the $N_f=1$ condensate.

I would like to thank my collaborators Roland Hoffmann, Zhaofeng Liu, and Stefan Schaefer for many
conversations, and I am grateful to
Adi Armoni,
Francesco Knechtli,
Francesco Sannino,
Misha Shifman,
Matt Strassler,
Mithat \"Unsal,
Gabriele Veneziano,
and Larry Yaffe
for correspondence.
And thanks to the organizers for putting on a fantastic conference!
This work was supported by the US Department of Energy.

\bibliographystyle{JHEP}
\bibliography{www}

\providecommand{\href}[2]{#2}\begingroup\raggedright\begin{thebibliography}{10}

\bibitem{Strassler:2001fs}
M.~J. Strassler, {\it On methods for extracting exact non-perturbative results
  in non-supersymmetric gauge theories},
  \href{http://xxx.lanl.gov/abs/hep-th/0104032}{{\tt hep-th/0104032}}.

\bibitem{Armoni:2003fb}
A.~Armoni, M.~Shifman, and G.~Veneziano, {\it Susy relics in one-flavor qcd
  from a new 1/n expansion},  {\em Phys. Rev. Lett.} {\bf 91} (2003) 191601,
  [\href{http://xxx.lanl.gov/abs/hep-th/0307097}{{\tt hep-th/0307097}}].

\bibitem{Armoni:2003gp}
A.~Armoni, M.~Shifman, and G.~Veneziano, {\it Exact results in
  non-supersymmetric large n orientifold field theories},  {\em Nucl. Phys.}
  {\bf B667} (2003) 170--182,
  [\href{http://xxx.lanl.gov/abs/hep-th/0302163}{{\tt hep-th/0302163}}].

\bibitem{Armoni:2003yv}
A.~Armoni, M.~Shifman, and G.~Veneziano, {\it Qcd quark condensate from susy
  and the orientifold large-n expansion},  {\em Phys. Lett.} {\bf B579} (2004)
  384--390, [\href{http://xxx.lanl.gov/abs/hep-th/0309013}{{\tt
  hep-th/0309013}}].

\bibitem{Armoni:2004ub}
A.~Armoni, M.~Shifman, and G.~Veneziano, {\it Refining the proof of planar
  equivalence},  {\em Phys. Rev.} {\bf D71} (2005) 045015,
  [\href{http://xxx.lanl.gov/abs/hep-th/0412203}{{\tt hep-th/0412203}}].

\bibitem{'tHooft:1974hx}
G.~'t~Hooft, {\it A two-dimensional model for mesons},  {\em Nucl. Phys.} {\bf
  B75} (1974) 461.

\bibitem{Armoni:2004uu}
A.~Armoni, M.~Shifman, and G.~Veneziano, {\it From super-yang-mills theory to
  qcd: Planar equivalence and its implications},
  \href{http://xxx.lanl.gov/abs/hep-th/0403071}{{\tt hep-th/0403071}}.

\bibitem{Patella:2005vx}
A.~Patella, {\it A insight on the proof of orientifold planar equivalence on
  the lattice},  {\em Phys. Rev.} {\bf D74} (2006) 034506,
  [\href{http://xxx.lanl.gov/abs/hep-lat/0511037}{{\tt hep-lat/0511037}}].

\bibitem{Unsal:2006pj}
M.~Unsal and L.~G. Yaffe, {\it (in)validity of large n orientifold
  equivalence},  \href{http://xxx.lanl.gov/abs/hep-th/0608180}{{\tt
  hep-th/0608180}}.

\bibitem{Davies:1999uw}
N.~M. Davies, T.~J. Hollowood, V.~V. Khoze, and M.~P. Mattis, {\it Gluino
  condensate and magnetic monopoles in supersymmetric gluodynamics},  {\em
  Nucl. Phys.} {\bf B559} (1999) 123--142,
  [\href{http://xxx.lanl.gov/abs/hep-th/9905015}{{\tt hep-th/9905015}}].

\bibitem{DeGrand:2006ws}
T.~DeGrand and S.~Schaefer, {\it Simulating an arbitrary number of flavors of
  dynamical overlap fermions},  {\em JHEP} {\bf 07} (2006) 020,
  [\href{http://xxx.lanl.gov/abs/hep-lat/0604015}{{\tt hep-lat/0604015}}].

\bibitem{DeGrand:2006uy}
T.~DeGrand, R.~Hoffmann, S.~Schaefer, and Z.~Liu, {\it Quark condensate in
  one-flavor qcd},  {\em Phys. Rev.} {\bf D74} (2006) 054501,
  [\href{http://xxx.lanl.gov/abs/hep-th/0605147}{{\tt hep-th/0605147}}].

\bibitem{Leutwyler:1992yt}
H.~Leutwyler and A.~Smilga, {\it Spectrum of {D}irac operator and role of
  winding number in {QCD}},  {\em Phys. Rev.} {\bf D46} (1992) 5607--5632.

\bibitem{Shuryak:1992pi}
E.~V. Shuryak and J.~J.~M. Verbaarschot, {\it Random matrix theory and spectral
  sum rules for the dirac operator in qcd},  {\em Nucl. Phys.} {\bf A560}
  (1993) 306--320, [\href{http://xxx.lanl.gov/abs/hep-th/9212088}{{\tt
  hep-th/9212088}}].

\bibitem{Verbaarschot:1993pm}
J.~J.~M. Verbaarschot and I.~Zahed, {\it Spectral density of the qcd dirac
  operator near zero virtuality},  {\em Phys. Rev. Lett.} {\bf 70} (1993)
  3852--3855, [\href{http://xxx.lanl.gov/abs/hep-th/9303012}{{\tt
  hep-th/9303012}}].

\bibitem{Verbaarschot:1994qf}
J.~J.~M. Verbaarschot, {\it The spectrum of the qcd dirac operator and chiral
  random matrix theory: The threefold way},  {\em Phys. Rev. Lett.} {\bf 72}
  (1994) 2531--2533, [\href{http://xxx.lanl.gov/abs/hep-th/9401059}{{\tt
  hep-th/9401059}}].

\bibitem{Damgaard:2000qt}
P.~H. Damgaard, U.~M. Heller, R.~Niclasen, and K.~Rummukainen, {\it Eigenvalue
  distributions of the qcd dirac operator},  {\em Phys. Lett.} {\bf B495}
  (2000) 263--270, [\href{http://xxx.lanl.gov/abs/hep-lat/0007041}{{\tt
  hep-lat/0007041}}].

\bibitem{Damgaard:2000ah}
P.~H. Damgaard and S.~M. Nishigaki, {\it Distribution of the k-th smallest
  dirac operator eigenvalue},  {\em Phys. Rev.} {\bf D63} (2001) 045012,
  [\href{http://xxx.lanl.gov/abs/hep-th/0006111}{{\tt hep-th/0006111}}].

\bibitem{Sommer:1993ce}
R.~Sommer, {\it A new way to set the energy scale in lattice gauge theories and
  its applications to the static force and alpha-s in su(2) yang-mills theory},
   {\em Nucl. Phys.} {\bf B411} (1994) 839--854,
  [\href{http://xxx.lanl.gov/abs/hep-lat/9310022}{{\tt hep-lat/9310022}}].

\bibitem{Martinelli:1994ty}
G.~Martinelli, C.~Pittori, C.~T. Sachrajda, M.~Testa, and A.~Vladikas, {\it A
  general method for nonperturbative renormalization of lattice operators},
  {\em Nucl. Phys.} {\bf B445} (1995) 81--108,
  [\href{http://xxx.lanl.gov/abs/hep-lat/9411010}{{\tt hep-lat/9411010}}].

\bibitem{Neuberger:1997fp}
H.~Neuberger, {\it Exactly massless quarks on the lattice},  {\em Phys. Lett.}
  {\bf B417} (1998) 141--144,
  [\href{http://xxx.lanl.gov/abs/hep-lat/9707022}{{\tt hep-lat/9707022}}].

\bibitem{Neuberger:1998my}
H.~Neuberger, {\it A practical implementation of the overlap-dirac operator},
  {\em Phys. Rev. Lett.} {\bf 81} (1998) 4060--4062,
  [\href{http://xxx.lanl.gov/abs/hep-lat/9806025}{{\tt hep-lat/9806025}}].

\bibitem{Ginsparg:1981bj}
P.~H. Ginsparg and K.~G. Wilson, {\it A remnant of chiral symmetry on the
  lattice},  {\em Phys. Rev.} {\bf D25} (1982) 2649.

\bibitem{Bode:1999dd}
A.~Bode, U.~M. Heller, R.~G. Edwards, and R.~Narayanan, {\it First experiences
  with hmc for dynamical overlap fermions},
  \href{http://xxx.lanl.gov/abs/hep-lat/9912043}{{\tt hep-lat/9912043}}.

\bibitem{Cundy:2005mr}
N.~Cundy, {\it Current status of dynamical overlap project},  {\em Nucl. Phys.
  Proc. Suppl.} {\bf 153} (2006) 54--61,
  [\href{http://xxx.lanl.gov/abs/hep-lat/0511047}{{\tt hep-lat/0511047}}].

\bibitem{Fodor:2003bh}
Z.~Fodor, S.~D. Katz, and K.~K. Szabo, {\it Dynamical overlap fermions, results
  with hybrid monte-carlo algorithm},  {\em JHEP} {\bf 08} (2004) 003,
  [\href{http://xxx.lanl.gov/abs/hep-lat/0311010}{{\tt hep-lat/0311010}}].

\bibitem{DeGrand:2004nq}
T.~DeGrand and S.~Schaefer, {\it Physics issues in simulations with dynamical
  overlap fermions},  {\em Phys. Rev.} {\bf D71} (2005) 034507,
  [\href{http://xxx.lanl.gov/abs/hep-lat/0412005}{{\tt hep-lat/0412005}}].

\bibitem{DeGrand:2005vb}
T.~A. DeGrand and S.~Schaefer, {\it Chiral properties of two-flavor qcd in
  small volume and at large lattice spacing},  {\em Phys. Rev.} {\bf D72}
  (2005) 054503, [\href{http://xxx.lanl.gov/abs/hep-lat/0506021}{{\tt
  hep-lat/0506021}}].

\bibitem{Morningstar:2003gk}
C.~Morningstar and M.~J. Peardon, {\it Analytic smearing of su(3) link
  variables in lattice qcd},  {\em Phys. Rev.} {\bf D69} (2004) 054501,
  [\href{http://xxx.lanl.gov/abs/hep-lat/0311018}{{\tt hep-lat/0311018}}].

\bibitem{DeGrand:2000tf}
{\bf MILC} Collaboration, T.~A. DeGrand, {\it A variant approach to the overlap
  action},  {\em Phys. Rev.} {\bf D63} (2001) 034503,
  [\href{http://xxx.lanl.gov/abs/hep-lat/0007046}{{\tt hep-lat/0007046}}].

\bibitem{Press:2002}
W.~H. Press, S.~A. Teukolsky, and B.~P.~F. William T.~Vetterling, {\em
  Numerical Recipes in C++ : The Art of Scientific Computing}.
\newblock Cambridge Univ. Press, 2nd~ed., 2002.

\bibitem{McNeile:2005pd}
C.~McNeile, {\it An estimate of the chiral condensate from unquenched lattice
  qcd},  {\em Phys. Lett.} {\bf B619} (2005) 124--128,
  [\href{http://xxx.lanl.gov/abs/hep-lat/0504006}{{\tt hep-lat/0504006}}].

\bibitem{DeGrand:2006nv}
T.~DeGrand, Z.~Liu, and S.~Schaefer, {\it Quark condensate in two-flavor qcd},
  \href{http://xxx.lanl.gov/abs/hep-lat/0608019}{{\tt hep-lat/0608019}}.

\bibitem{Gasser:1986vb}
J.~Gasser and H.~Leutwyler, {\it Light quarks at low temperatures},  {\em Phys.
  Lett.} {\bf B184} (1987) 83.

\bibitem{Capitani:1998mq}
{\bf ALPHA} Collaboration, S.~Capitani, M.~Luscher, R.~Sommer, and H.~Wittig,
  {\it Non-perturbative quark mass renormalization in quenched lattice qcd},
  {\em Nucl. Phys.} {\bf B544} (1999) 669--698,
  [\href{http://xxx.lanl.gov/abs/hep-lat/9810063}{{\tt hep-lat/9810063}}].

\bibitem{DellaMorte:2004bc}
{\bf ALPHA} Collaboration, M.~Della~Morte {\em et~al.}, {\it Computation of the
  strong coupling in qcd with two dynamical flavours},  {\em Nucl. Phys.} {\bf
  B713} (2005) 378--406, [\href{http://xxx.lanl.gov/abs/hep-lat/0411025}{{\tt
  hep-lat/0411025}}].

\bibitem{Armoni:2005wt}
A.~Armoni, G.~Shore, and G.~Veneziano, {\it Quark condensate in massless qcd
  from planar equivalence},  {\em Nucl. Phys.} {\bf B740} (2006) 23--35,
  [\href{http://xxx.lanl.gov/abs/hep-ph/0511143}{{\tt hep-ph/0511143}}].

\bibitem{Sannino:2003xe}
F.~Sannino and M.~Shifman, {\it Effective lagrangians for orientifold
  theories},  {\em Phys. Rev.} {\bf D69} (2004) 125004,
  [\href{http://xxx.lanl.gov/abs/hep-th/0309252}{{\tt hep-th/0309252}}].

\bibitem{Gorsky:2004nb}
A.~Gorsky and M.~Shifman, {\it Spectral degeneracy in supersymmetric
  gluodynamics and one- flavor qcd related to n = 1/2 susy},  {\em Phys. Rev.}
  {\bf D71} (2005) 025009, [\href{http://xxx.lanl.gov/abs/hep-th/0410099}{{\tt
  hep-th/0410099}}].

\bibitem{Dashen:1970et}
R.~F. Dashen, {\it Some features of chiral symmetry breaking},  {\em Phys.
  Rev.} {\bf D3} (1971) 1879--1889.

\bibitem{Smilga:1998dh}
A.~V. Smilga, {\it {QCD} at theta approx. pi},  {\em Phys. Rev.} {\bf D59}
  (1999) 114021, [\href{http://xxx.lanl.gov/abs/hep-ph/9805214}{{\tt
  hep-ph/9805214}}].

\bibitem{Creutz:2000bs}
M.~Creutz, {\it Aspects of chiral symmetry and the lattice},  {\em Rev. Mod.
  Phys.} {\bf 73} (2001) 119--150,
  [\href{http://xxx.lanl.gov/abs/hep-lat/0007032}{{\tt hep-lat/0007032}}].

\bibitem{Durr:2006ze}
S.~Durr and C.~Hoelbling, {\it Lattice fermions with complex mass},  {\em Phys.
  Rev.} {\bf D74} (2006) 014513,
  [\href{http://xxx.lanl.gov/abs/hep-lat/0604005}{{\tt hep-lat/0604005}}].

\end{thebibliography}\endgroup

\end{document}